\documentclass[12pt,preprint]{aastex}
\usepackage{natbib}
\usepackage{amsmath}
\usepackage{graphicx}

\def\aap{A\&A\,  }
\def\apj{ApJ\,  }
\def\apjl{ApJ\,  }
\def\apjs{ApJS  }

\def\mnras{MNRAS\,  }
\def\pasp{PASP  }

\def\sun{\hbox{$\odot$}}
\def\sunmass{M_{\sun}}
\def\2F1{~_2F_1}
\begin{document}
\shorttitle
{
IMF and truncated beta
}
\shortauthors{Zaninetti}
\title
{
The initial mass function modeled
by a left truncated beta distribution
}
\author{Lorenzo Zaninetti  }

\affil {Dipartimento di Fisica,
        Via Pietro Giuria 1,\\
        10125 Torino, Italy}

\email   {zaninetti@ph.unito.it  \\
\url     {http://www.ph.unito.it/$\tilde{~}$zaninett}}
\begin{abstract}
The initial  mass  function (IMF) for the stars
is usually  fitted by  three
straight lines, which  means  seven parameters.
The  presence of brown dwarfs (BD) increases
to four  the straight lines and to nine
the parameters.
Another common fitting function is the lognormal
distribution, which is characterized
by two parameters.
This paper is devoted  to demonstrating  the advantage
of introducing  a left truncated beta probability
density function, which is characterized
by four parameters.
The constant of normalization,
the  mean,
the mode and  the distribution function are calculated
for  the left truncated beta distribution.
The normal-beta (NB) distribution
which results from convolving
independent  normally  distributed and beta distributed
components is also derived.
The chi-square  test and the K-S test are  performed
on a first sample  of stars  and BDs
which belongs to  the massive young
cluster  NGC 6611
and on a second sample
which represents the star's  masses of the  cluster NGC 2362.
\end{abstract}

\section{Introduction}

The distribution in mass of the stars
has  been fitted with a power law
starting with \cite{Salpeter1955}.
He suggested
$\xi ( {{m}}) \propto   {{m}}^{-\alpha}$
where $\xi   ( {{m}})$ represents  the probability
of having  a mass between $ {{m}}$ and
$ {{m}}+d{{m}}$;
He found  $\alpha= 2.35$
in the range $10  {M}_{\sun}~>~ {M} \geq  1  {M}_{\sun}$;
this value
 has changed  little
 with time and a recent evaluation quotes 2.3, see~\cite{Kroupa2001}.
Subsequent research has started to analyze the initial  mass
function (IMF) with three  power laws, see
\cite{Scalo1986,Kroupa1993,Binney1998} and four power laws, see
\cite{Kroupa2001}. A first comment  on this temporal evolution is
that the  name is not appropriate because the  power  function
distribution $\xi ( {{m}}) \propto   {{m}}^{b}$ is defined only
for positive values of $b$, see \cite{evans}. A second comment  is
that the  exact name for a probability density function (PDF) $\xi
( {{m}}) \propto   {{m}}^{-(c+1)}$ with $c>0$ is the Pareto
distribution. A third comment is that this  progressive increase
in the number of segments has limited the development  of new or
modified PDFs. The  approach to the IMF by a continuous
distribution has been modeled by the lognormal distribution in
order to fit both the range of the stars and  the brown dwarfs
(BDs) regime, see \cite{Chabrier2003}. Recall that usually  the
standard PDFs such as the lognormal, gamma, generalized gamma, and
Weibull are defined  in the interval $0 \leq x <\infty$. The fact
that the number of stars with  mass $ m < 0.07 \sunmass $ is
nearly  zero  suggests   a left truncated PDF. Our  analysis  has
therefore been focused on the beta distribution, which by
definition has an upper bound. From the previous  analysis the
following questions can be  raised.
\begin{itemize}
\item
Is it possible to find  the constant  of normalization
for  a left  truncated beta PDF?
\item
Is it possible to derive an analytical  expression
for the mean, the mode, and the distribution function
of a left  truncated beta PDF?
\item
Is  a left  truncated beta PDF an acceptable model
for the IMF as well as a real  sample  of masses?
\end{itemize}
In order to  answer the previous  questions,
we first  review  some standard PDFs,
see  Section \ref{existing}.
We subsequently  introduce  the various
beta PDFs,
the convolution of a beta PDF with a normal  PDF,
 and a left truncated
beta,
see Section \ref{secbeta}.
In order to find out  which PDF performs best,
the two main criteria which report
the  goodness  of fit are  found
in Section \ref{goodness}.
A comparison between various continuous PDFs
and the left truncated
beta is carried out in Sections  \ref{secngc2362}
and \ref{secngc6611}
for two samples of stars.

\section{Distributions commonly used}

This  section reviews  some  standard  PDFs,
namely, the lognormal, gamma,
generalized gamma, Pareto,
truncated Pareto, and   the recently developed
Double Pareto-lognormal distribution.

\label{existing}

\subsection{Lognormal distribution}

Let $X$ be a random variable taking
values $x$ in the interval
$[0, \infty]$; the {\em lognormal} PDF
, following \cite{evans}
or formula (14.2)$^\prime$ in
\cite{univariate1},  is
\begin{equation}
f_{LN} = \frac
{
1
}
{
x \sigma \sqrt {2 \pi}
}
\exp { \frac {-\left [ \ln (x/m) \right ] ^2}{2\sigma^2} }
\quad,
\label{pdflognormal1}
\end{equation}
 or
\begin{equation}
f_{LN} = \frac
{
1
}
{
x \sigma \sqrt {2 \pi}
}
\exp { \frac {-\left ( \ln x - \mu_{LN} \right ) ^2}{2\sigma^2} }
\quad,
\label{pdflognormal2}
\end{equation}
where  $m=\exp{\mu_{LN}}$ and
$\mu_{LN}=\log {m}$.

\subsection{Gamma distribution}

Let $X$ be a random variable taking
values $x$ in the interval
$[0, \infty]$;
the {\em gamma} PDF is
\begin{equation}
p(x;b,c) =
\frac
{
\left( {\frac {x}{b}} \right) ^{c-1}{{\rm e}^{-{\frac {x}{b}}}}
}
{
b \mathop{\Gamma}  \left( c \right)
}
\quad,
\label{pdfgamma}
\end{equation}
where $\mathop{\Gamma}(z)$ is the gamma function
\begin{equation}
\mathop{\Gamma\/}\nolimits\!\left(z\right)=\int
_{0}^{\infty}e^{{-t}}t^{{z-1}}dt \quad,
\end{equation}
see formula (17.1) in \cite{univariate1}.

\subsection{Generalized  gamma distribution}

Let $X$ be a random variable taking values
$x$ in the interval
$[0, \infty]$; the {\em generalized gamma} PDF,
following \cite{evans},
is
\begin{equation}
f(x;a,b,c) = c \frac {b^{a/c}} {\Gamma (a/c) } x^{a-1} \exp{(-b
x^c)} \quad, \label{pdfgammageneralized}
\end{equation}
see formula (17.116) in \cite{univariate1}.

\subsection{The Pareto and the truncated  Pareto distributions}

Let $X$ be a random variable taking values $x$ in the interval
$[a, \infty]$, $a>0$.
The  {\em Pareto} PDF
is defined by
\begin {equation}
f(x;a,c) = {c a^c}{x^{-(c+1)}} \quad,
\label{pdfpareto}
\end {equation}
with $ c~>0$, see formula (20.3) in \cite{univariate1}.
The  traditional Salpeter
slope  is  therefore -(c+1).
An upper truncated Pareto random variable is defined
in the interval
$[a,b]$ and the corresponding PDF, following

\cite{Goldstein2004,Aban2006,Zaninetti2008d,White2008},
is
\begin {equation}
f_T(x;a,b,c ) = \frac{ca^cx^{-(c+1)}}{1-\left (\frac{a}{b}\right)^c}
\quad.
\label{paretotruncated}
\end {equation}
Their means   are
\begin{equation}
E(x;a,c) = {\frac {ac}{c-1}}
\end{equation}
and
\begin{equation}
E(x;a,b,c)_T = \frac { ca \left( -1+ \left( {\frac {a}{b}} \right)
^{c-1} \right) } { \left( c-1 \right)  \left( -1+ \left( {\frac
{a}{b}} \right) ^{c}
 \right)
}
\quad.
\end{equation}

\subsection{The double Pareto-lognormal distribution }

The double Pareto lognormal  distribution has
been recently derived,
see formula (22) in
\cite{Reed2004}, and
 has been used to fit the actual sizes of cities
, see \cite{Giesen2010}
;  its PDF is
\begin{eqnarray}
f(x;\alpha,\beta,\mu,\sigma) =
1/2\,\alpha\,\beta\,( {{\rm e}^{1/2\,\alpha\,( \alpha\,{
\sigma}^{2}+2\,\mu-2\,\ln ( x )  ) }}{\it erfc}
( 1/2\,{\frac {( \alpha\,{\sigma}^{2}+\mu-\ln ( x
 )  ) \sqrt {2}}{\sigma}} )
\nonumber \\
+{{\rm e}^{1/2\,\beta\,
( \beta\,{\sigma}^{2}-2\,\mu+2\,\ln ( x )  ) }
}{\it erfc}( 1/2\,{\frac {( \beta\,{\sigma}^{2}-\mu+\ln
( x )  ) \sqrt {2}}{\sigma}} )  ) {x}^{-
1}( \alpha+\beta ) ^{-1},
\end{eqnarray}
where $\alpha$ and $\beta$
are the Pareto coefficients for the
upper and the lower tail,
respectively,
$\mu $ and $\sigma$  are
the lognormal body parameters, and
$erfc$  is
the complementary error function.
The parameters can be found minimizing 
the maximum  distance, $D$, 
of the K-S test, see Sect.  \ref{goodness}.

\section{Various Beta distributions}

This section reviews the  beta PDF
defined in $[0, 1]$, the  beta with scale  PDF
defined in $[0, b]$  and  the  general beta
defined in $[a, b]$.
The left  truncated   beta PDF
defined in $[a, b]$ but with a finite value
of probability at $x=a$  is explored.
The convolution of a beta distribution  with a normal distribution
is  also discussed.

\label{secbeta}

\subsection{Beta distribution}

Let $X$ be a random variable taking values $x$ in the interval
$[0, 1]$; the {\em beta} PDF  is
\begin {equation}
f(x;\alpha,\beta) = \frac{{x}^{\alpha-1} \left( 1-x \right)
^{\beta-1}}{\mathrm{B}  \left( \alpha,\beta \right) } \quad,
\label{betastandard}
\end {equation}
with  $\alpha>0$ and $\beta>0$, see \cite{evans}
or  formula (25.2) in  \cite{univariate2}.
Here $\mathrm{B}$
is the beta function defined by
\begin{equation}
\mathop{\mathrm{B}\/}\nolimits\!\left(a,b\right)=\int
_{0}^{1}t^{{a-1}}(1-t)^{{b-1}}dt=\frac{\mathop{\Gamma\/}\nolimits\!\left(a\right)\mathop{\Gamma\/}\nolimits\!\left(b\right)}{\mathop{\Gamma\/}\nolimits\!\left(a+b\right)}
\quad.
\end{equation}
Its mean   is
\begin{equation}
E(x;\alpha,\beta)=  {\frac {\alpha}{\alpha+\beta}} \quad,
\end{equation}
and its variance,
\begin{equation}
\sigma^2(x;\alpha,\beta) = {\frac {\alpha\,\beta}{ \left(
1+\alpha+\beta \right) \left( \alpha+ \beta \right) ^{2}}} \quad,
\end{equation}
see  formula (25.15a) in \cite{univariate2}.
The mode  is at
\begin{equation}
m(x;\alpha,\beta)={\frac {\alpha-1}{\alpha-2+\beta}} \quad.
\end{equation}
The  method of matching moments gives
the following
parameter
estimation
\begin{equation}
\tilde{\alpha}= {\it \bar{x}}\, \left( {\frac {{\it \bar{x}}\,
\left( 1-{\it \bar{x}}
 \right) }{{s}^{2}}}-1 \right)
 \quad,
 \end{equation}
and
\begin{equation}
\tilde{\beta}=
 \left( 1-{\it \bar{x}} \right)  \left( {\frac {{\it \bar{x}}\, \left( 1
-{\it \bar{x}} \right) }{{s}^{2}}}-1 \right) \quad,
\end{equation}
where $\bar{x}$ and  $s^2$ are the mean and  the
variance of the
sample.
The distribution function   (DF)  is
\begin{equation}
DF(x;\alpha,\beta) =
\frac{{x}^{\alpha}{\mbox{$_2$F$_1$}(\alpha,-\beta+1;\,\alpha+1;\,x)}}
{\beta  \left( \alpha,\beta \right) \alpha} \quad,
\end{equation}
where ${\2F1(a,b;\,c;\,z)}$ is the
regularized hypergeometric
function
\cite{Abramowitz1965,Seggern1992,Thompson1997,Gradshteyn2007,NIST2010}.

\subsection{Beta distribution with scale}

Let $X$ be a random variable taking
values $x$ in the interval
$[0, b]$.
The {\em beta
with scale} PDF  is
\begin {equation}
f_b(x;b,\alpha,\beta) = \frac{\left( {\frac {x}{b}} \right)
^{\alpha-1} \left( 1-{\frac {x}{b}}
 \right) ^{\beta-1}} {\mathrm{B} \left( \alpha,\beta \right) b}.
\label{pdfbetab}
\end {equation}

Its expected mean is
\begin{equation}
E(x;b,\alpha,\beta) = {\frac {\alpha\,b}{\alpha+\beta}}  \quad,
\end{equation}
and its variance,
\begin{equation}
\sigma(x;b,\alpha,\beta)_{b}^2 ={\frac {\beta\,\alpha\,{b}^{2}}{
\left( 1+\alpha+\beta \right)
 \left( \alpha+\beta \right) ^{2}}}
 \quad.
\end{equation}
The mode  is at
\begin{equation}
m(x;b,\alpha,\beta)_b= {\frac {b \left( \alpha-1 \right)
}{\beta-2+\alpha}}\quad.
\end{equation}
The  DF  is
\begin{equation}
DF_b(x;b,\alpha,\beta) =
\frac{{x}^{\alpha}{\mbox{$_2$F$_1$}(\alpha,-\beta+1;\,\alpha+1;\,x)}}{
\beta  \left( \alpha,\beta \right) \alpha}
 \quad.
\end{equation}
The three  parameters can be estimated by
\begin{equation}
\tilde{b}= maximum ~of~sample
 \quad,
 \end{equation}

\begin{equation}
\tilde{\alpha}=-{\frac {{\it \bar{x}}\, \left( -b{\it
\bar{x}}+{{\it \bar{x}}}^{2}+{s}^{ 2} \right) }{{s}^{2}b}}
  \quad,
  \label{alphavalue}
\end{equation}

\begin{equation}
\tilde{\beta}= -{\frac { \left( b-{\it \bar{x}} \right)  \left(
-b{\it \bar{x}}+{{\it \bar{x}}}^{2}+{s}^{2} \right) }{{s}^{2}b}}
  \quad.
  \label{betavalue}
\end{equation}

\subsection{General Beta distribution}

Let $X$ be a random variable taking values $x$ in
the interval
$[a, b]$. The {\em general beta} PDF
is
\begin {equation}
f_{ab}(x;a,b,\alpha,\beta)= \frac{\left( b-a \right)  \left( x-a
\right) ^{\alpha-1} \left( b-x
 \right) ^{\beta-1}}
 {{b}^{\alpha+\beta-1}b \left( {\frac {b-a}{b}} \right) ^{\alpha+\beta}
\mathrm{B}  \left( \alpha,\beta \right) } \quad,
\label{pdfbetaab}
\end{equation}
see formula (25.1) in \cite{univariate2}.
Its expected mean is
\begin{equation}
E(x;a,b,\alpha,\beta)_{ab} =  {\frac
{\alpha\,b+a\beta}{\alpha+\beta}}\quad,
\end{equation}
and its variance,
\begin{equation}
\sigma(x;a,b,\alpha,\beta)_{ab}^2 = {\frac { \left( a-b \right)
^{2}\alpha\,\beta}{ \left( \alpha+\beta+1
 \right)  \left( \alpha+\beta \right) ^{2}}}
 \quad.
\end{equation}
The mode  is at
\begin{equation}
m(x;a,b,\alpha,\beta)_{ab}=  {\frac
{\alpha\,b-b+a\beta-a}{-2+\alpha+\beta}}\quad.
\end{equation}

The four parameters can be estimated by
\begin{equation}
\tilde{a}= minimum ~of~sample \quad \tilde{b}= maximum ~of~sample
 \quad,
 \end{equation}

\begin{equation}
\tilde{\alpha}=-{\frac { \left( -{\it \bar{x}}+{\it \tilde{a}}
\right) \left( -{\it \bar{x}}\,{\it \tilde{b} }+{\it
\tilde{b}}\,{\it \tilde{a}}+{{\it \bar{x}}}^{2}+{s}^{2}-{\it
\tilde{a}}\,{\it \bar{x}}
 \right) }{{s}^{2} \left( {\it \tilde{a}}-{\it \tilde{b}} \right) }}
  \quad,
\end{equation}

\begin{equation}
\tilde{\beta}={\frac { \left( {\it \tilde{b}}-{\it \bar{x}}
\right) \left( -{\it \bar{x}}\,{\it \tilde{b}}+ {\it
\tilde{b}}\,{\it \tilde{a}}+{{\it \bar{x}}}^{2}+{s}^{2}-{\it
\tilde{a}}\,{\it \bar{x}} \right) } {{s}^{2} \left( {\it
\tilde{a}}-{\it \tilde{b}} \right) }}
  \quad,
\end{equation}
see  the discussion in section 25.4  of \cite{univariate2}.

\subsection{Left truncated beta distribution with scale}

\label{sectruncated}
Let $X$ be a random variable taking values $x$
in the interval
$[a, b]$ and having a finite value in $a$. The
{\em left truncated beta with scale} PDF is
\begin {equation}
f_T(x;a,b,\alpha,\beta) = K\,{x}^{\alpha-1} \left( b-x \right)
^{\beta-1},
\label{betatruncated}
\end{equation}
where the constant is
\begin{equation}
K=\frac{-\alpha\,\Gamma  \left( \alpha+\beta \right)}
{{b}^{\beta-1} H\,{a}^{ \alpha}\Gamma  \left( \alpha+\beta \right)
-{b}^{\beta-1+\alpha} \Gamma  \left( 1+\alpha \right) \Gamma
\left( \beta \right)},
\end{equation}
and
\begin{equation}
 H={\mbox{$_2$F$_1$}(\alpha,-\beta+1;\,1+\alpha;\,{\frac
{a}{b}})} \quad.
\end{equation}
The constant of normalization can  be obtained
from the integral  of the
beta  with scale  PDF
as represented by Equation (\ref{pdfbetab}).
The
integral 3.194.1 on p\. 315 of \cite {Gradshteyn2007},
\begin {eqnarray}
\int _{0}^{u}\!{x}^{\mu-1} \left( 1+\beta\,x \right) ^{\nu}{dx}
\nonumber  \\
=
{\frac {{{\rm e}^{\mu\,\ln  \left( u \right) }}
{\mbox{$_2$F$_1$}(\mu,-\nu;\,1+\mu;\,-\beta\,u)}}{\mu}},
\end{eqnarray}
is useful in the analytical  derivation of the constant.
This  PDF,
which is new and therefore cannot  be found
in \cite{univariate2},
at  $x=a$
is not zero but takes the finite value
\begin {equation}
f_T(a;a,b,\alpha,\beta) =K\,{a}^{\alpha-1} \left( b-a \right)
^{\beta-1}
 \quad.
\label{pdfbetabtruncated}
\end{equation}
Its expected mean is
\begin{eqnarray}
E(x;a,b,\alpha,\beta)_T =K\,({\frac {{b}^{\beta-1}{b}^{1+\alpha}
{\mbox{$_2$F$_1$}(-\beta+1,1+\alpha;\,2+\alpha;\,1)}}{1+\alpha}}
\nonumber \\
\frac{-{b}^{\beta-1}{a}^{1+\alpha}
{\mbox{$_2$F$_1$}(-\beta+1,1+\alpha;\,2+\alpha;\,{\frac {a}{b}})}
} { 1+\alpha}
 )
 \quad.
 \label{meantruncated}
 \end{eqnarray}
The mode  is at
\begin{equation}
m(x;b,\alpha,\beta)_T={\frac {b \left( \alpha-1 \right)
}{\alpha-2+\beta}} \quad,
\end{equation}
and in order to exist one must have $m_T > a$.
The
 DF  is
\begin{eqnarray}
DF_T(x;a,b,\alpha,\beta) = \nonumber \\ \Gamma  \left(
\alpha+\beta \right) {x}^{\alpha}
{\mbox{$_2$F$_1$}(\alpha,-\beta+1;\,1+\alpha;\,{\frac
{x}{b}})}{{\it H_D}}^{-1} \nonumber \\ -
{\mbox{$_2$F$_1$}(\alpha,-\beta+1;\,1+\alpha;\,{\frac
{a}{b}})}{a}^{ \alpha}\Gamma  \left( \alpha+\beta \right) {{\it
H_D}}^{-1}
\label{betatruncatedf}
 \quad,
\end{eqnarray}
where
\begin{eqnarray}
H_D= \nonumber
\\{\mbox{$_2$F$_1$}(\alpha,-\beta+1;\,1+\alpha;\,{\frac
{a}{b}})}{a}^{ \alpha}\Gamma  \left( \alpha+\beta \right) -\Gamma
\left( \beta
 \right) {b}^{\alpha}\Gamma  \left( \alpha \right) \alpha
\quad.
\end{eqnarray}
The survival  function is
\begin{equation}
S_T(x;a,b,\alpha,\beta) = 1 - DF_T(x;a,b,\alpha,\beta)
\label{betatruncatedsurv}
\quad.
\end{equation}
The four parameters can be estimated by
\begin{equation}
\tilde{a}= minimum ~of~sample \quad \tilde{b}= maximum ~of~sample
 \quad.
 \end{equation}
A first  couple for  $\tilde{\alpha}$ and
$\tilde{\beta}$ can be
obtained from those of the beta distribution with scale as given
by Equations (\ref{alphavalue}) and (\ref{betavalue}).
A subsequent
numerical loop  around the previous values gives the couple
which
minimize the $\chi^2$.

\subsection{Beta distribution + normal}

We consider the  sum $Z=X+Y$  where $X$  is a standard
normal random variable, $N(x;\sigma)$, and $Y$ is a
general beta distribution,
$f_{ab}(y;a,b,\alpha,\beta)$,
as  represented by the PDF (\ref{pdfbetaab}).
The sum, $NB(z;a,b,\alpha,\beta,\sigma)$,  is
\begin{equation}
NB(z;a,b,\alpha,\beta,\sigma)=
\int_a^b N(z-y)  f_{ab}(y;a,b,\alpha,\beta) dy
\quad.
\label{convolution}
\end{equation}
A  similar  example, uniform + normal, can be found  in
\citep[] [Sec. 5.11.2] {Brandt1998}.

This integral  has  an analytical solution for
$\alpha$ and $\beta$ integers,
for example, when $\alpha$=1  and $\beta$=1,
\begin{eqnarray}
NB(z;a,b,1,1,\sigma)=
\nonumber  \\
1/2\,
{{\rm erf}\left(1/2\,{\frac {\sqrt {2}a}{\sigma}}-1/2\,{\frac {\sqrt
{2}z}{\sigma}}\right)}
 \left( a-b \right) ^{-1}-1/2\,
{{\rm erf}\left(1/2\,{\frac {\sqrt {2}b}{\sigma}}-1/2\,{\frac {\sqrt
{2}z}{\sigma}}\right)}
 \left( a-b \right) ^{-1}
\quad,
\end{eqnarray}
where ${\rm erf}$ is the error function.

\section{Goodness of fit tests}

\label{goodness}

The occasional reader may question which  is
the best fit  for the distributions analyzed here.
In order  to answer  this question,
we first  introduce $\chi^2$, which
is computed
according to the formula
\begin{equation}
\chi^2 = \sum_{i=1}^n \frac { (T_i - O_i)^2} {T_i},
\label{chisquare}
\end {equation}
where $n  $   is the number of bins,
      $T_i$   is the theoretical value,
and   $O_i$   is the experimental value represented
by the frequencies.
The theoretical  frequency distribution
is given by:
\begin{equation}
 T_i  = N {\Delta x_i } p(x) \quad,
\label{frequenciesteo}
\end{equation}
where $N$ is the number of elements of the sample,
      $\Delta x_i $ is the magnitude of the size interval,
and   $p(x)$ is the PDF  under examination.
The size of the bins, $\Delta x_i $,
is equal for each bin
in the the case of linear histograms,
but different for each bin when
logarithmic  histograms  are considered.

A reduced  merit function $\chi_{red}^2$
is  evaluated  by
\begin{equation}
\chi_{red}^2 = \chi^2/NF
\quad,
\label{chisquarereduced}
\end{equation}
where $NF=n-k$ is the number of degrees  of freedom,
$n$ is the number of bins,
and $k$ is the number of parameters.
The goodness  of the fit can be expressed by
the probability $Q$, see  equation 15.2.12  in \cite{press},
which involves the degrees of freedom
and the $\chi^2$.
According to  \cite{press}, the
fit ``may be acceptable'' if  $Q>0.001$.
The Akaike information criterion
(AIC), see \cite{Akaike1974},
is defined by
\begin{equation}
AIC  = 2k - 2  ln(L)
\quad,
\end {equation}
where $L$ is
the likelihood  function  and $k$  the number of  free parameters
in the model.
We assume  a Gaussian distribution for  the errors
and  the likelihood  function
can be derived  from the $\chi^2$ statistic
$L \propto \exp (- \frac{\chi^2}{2} ) $
where  $\chi^2$ has been computed by
Equation~(\ref{chisquare}),
see~\cite{Liddle2004}, \cite{Godlowski2005}.
Now the AIC becomes
\begin{equation}
AIC  = 2k + \chi^2
\quad.
\label{AIC}
\end {equation}
We  also  perform
the Kolmogorov--Smirnov test (K-S),
see\cite{Kolmogoroff1941,Smirnov1948,Massey1951},
which does not  require binning the data.
The K-S test,
as implemented by the FORTRAN subroutine KSONE in \cite{press},
finds
the maximum  distance, $D$, between the theoretical
and the astronomical  DF
as well the  significance  level  $P_{KS}$ ,
see formulas  14.3.5 and 14.3.9  in \cite{press}.
Values of $ P_{KS} \geq 0.1 $
assures that the fit is acceptable.

\section{Astrophysical applications}

\label {seccomparison}

This section reviews the galactic IMF as modeled
by three and four power laws
PDFs and fits
the masses of the  cluster NGC 2362 and  the
cluster  NGC 6611 with the various
PDFs here considered.

\subsection{Galactic IMF}

The IMF  is usually modeled by two or three
power laws of the
type
\begin{equation}
p_{stars}(m) \propto x^{-\alpha_i} \quad,
\end {equation}
each zone  being  characterized  by a different
exponent  ${\alpha_i}$.
In order to have a PDF normalized  to unity, one must have
\begin{equation}
\sum _{i=1,3}  \int_{m_i}^{m_{i+1}} c_i m^{-\alpha_i} dm =1
\quad.
\label{uno}
\end{equation}
For example,  we start with $c_1$=1:
$c_2$   will be determined by
the following
equation
\begin{equation}
c_1 (0.5 - \epsilon)^{-\alpha_1} =
c_2 (0.5 + \epsilon)^{-\alpha_2}
\quad,
\end{equation}
where $\epsilon$ is a small number, e.g.,
$\epsilon =10^{-4}$.
In the previous equation we insert
$\alpha_1=1.3$  and  $\alpha_2=2.3$
and therefore $c_2$ = 0.503.
The same procedure  applied to $c_3$ gives
$c_3$= 0.506.
The integral of $p_{stars}(m)$ over the field of existence
now gives  4.14, but  according to
the requirement of normalization  as given
by Equation (\ref{uno}), it should be 1.
As a consequence, the  three constants are now
$c_1 =0.24 $,  $c_2 =0.1205 $, and  $c_3 =0.1206 $,
which   is the same as
equation (59) in \cite{Kroupa2012}
\begin{equation}
  p(m) =
   \begin{cases}
0.24\,{x}^{- 1.3}& \text{if } 0.07 \sunmass<m \leq 0.5 \sunmass\\
0.12\,{x}^{- 2.3}& \text{if } 0.5  \sunmass<m \leq 1.0 \sunmass\\
0.12\,{x}^{- 2.7}& \text{if } 1.0  \sunmass<m \leq 10 \sunmass
\quad.
   \end{cases}
\label{threepowerlaws}
\end{equation}
The  mean   of the  galactic IMF
is  given by
a numerical integration over
the three  zones
\begin{equation}
\bar{m}=\sum _{i=1,3}  \int_{m_i}^{m_{i+1}} c_i m \,m^{-\alpha_i} dm =
0.389 \sunmass
 \quad.
 \label{meantrealfa}
\end{equation}
The  presence  of the brown dwarfs
means  the use of four  power laws instead of three
power laws:
\begin{equation}
  p(m) =
   \begin{cases}
2.194\,{x}^{- 0.3}& \text{if } 0.01 \sunmass<m \leq 0.07 \sunmass\\
0.153\,{x}^{- 1.3}& \text{if } 0.07 \sunmass<m \leq 0.5 \sunmass\\
0.076\,{x}^{- 2.3}& \text{if } 0.5  \sunmass<m \leq 1.0 \sunmass\\
0.076\,{x}^{- 2.7}& \text{if } 1.0  \sunmass<m \leq 10 \sunmass
\quad,
   \end{cases}
\label {fourpowerlaws}
\end{equation}
where in order to have a continuous
PDF, the BDs  have the range  $  0.01 \sunmass <m
\leq 0.07 \sunmass $ rather than   $  0.01 \sunmass <m
\leq 0.15 \sunmass $,
see  equation (59) in \cite{Kroupa2012}.
We have covered the galactic four  power laws,
we now introduce  the generalized  four power laws
$p_G(m;-\alpha_1, -\alpha_2,-\alpha_3 ,-\alpha_4,m_1, 
m_2,m_3,m_4,m_5)$  which  in   the case  
of NGC 2362 is
\begin{equation}
p_G (m;  -0.01,
         -0.02,
         -1.1 ,
         -2.7 , 
          0.01,
          0.07, 
          0.50, 
          1.0 , 
          10.) 
\quad   NGC~2362~case
\quad ,
\label{fourpowerngc2362}
\end{equation}
and in the case 
of NGC 6611 is
\begin{equation}
p_G (m; -0.01,
        -0.6,
        -2.4,
        -2.7, 
        0.01,
        0.07, 
        0.50, 
        1.0 , 
        10.) 
\quad   NGC~6611~case
\quad .
\label{fourpowerngc6611}
\end{equation}

\subsection{IMF of NGC 2362}
\label{secngc2362}
A photometric survey  of  NGC 2362  allows of deducing
the mass of 271 stars
in  the  range
 $1.47  {M}_{\sun}~>~ {M} \geq  0.11  {M}_{\sun}$,
see  \cite{Irwin2008} and the data
in
 J/MNRAS/384/675 at
the Centre de Donns astronomiques de Strasbourg (CDS). Table
\ref{chi2valuesngc2362}  shows the values of  $\chi_{red}^2$, the
AIC,  the probability $Q$, of  the  astrophysical   fits and the
results of the K-S test.

\begin{figure*}
\begin{center}
\includegraphics[width=10cm]{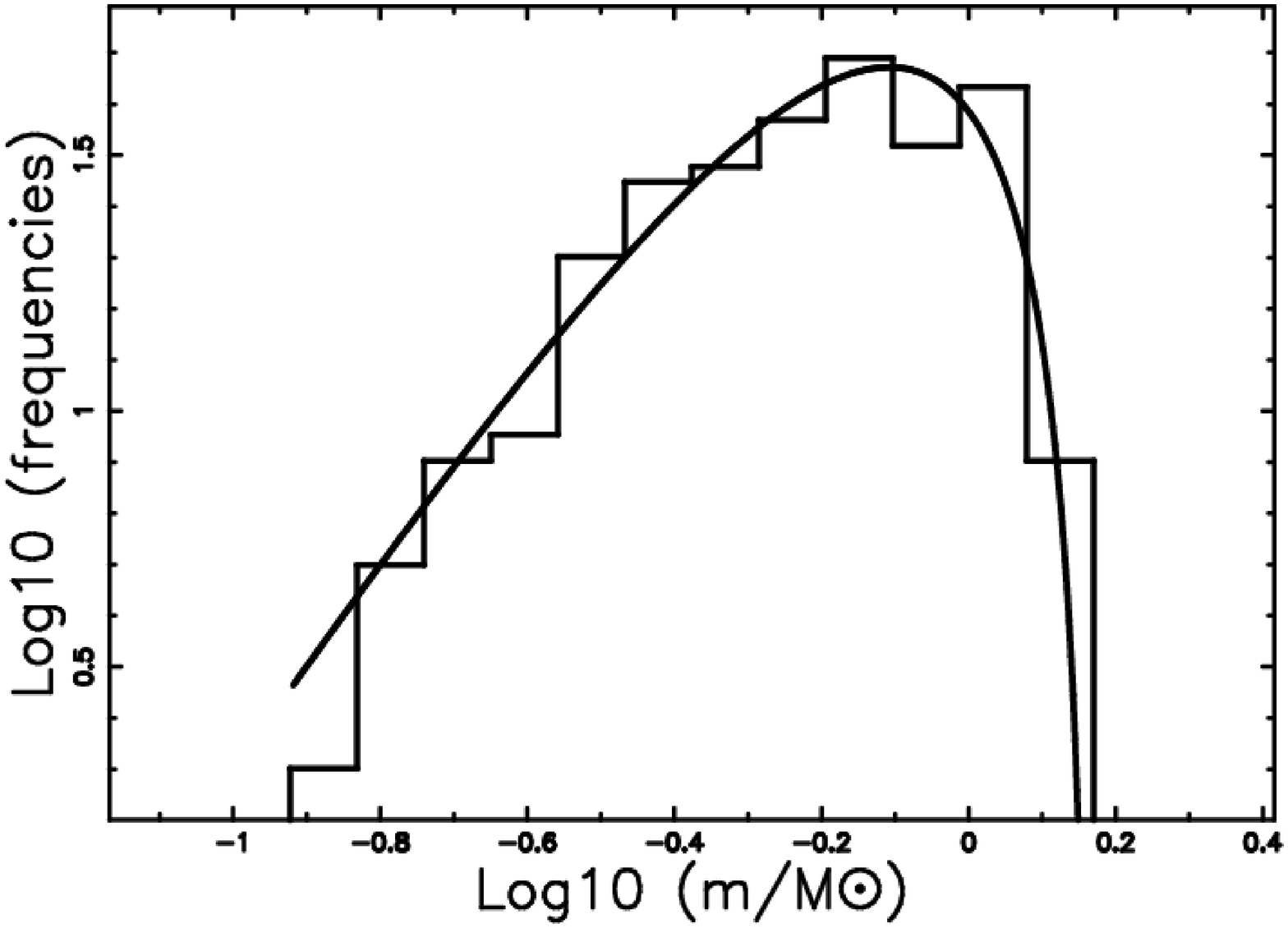}
\end{center}
\caption
{
Logarithmic histogram   of  mass distribution
as  given by   NGC 2362 cluster data (272 stars)
with a superposition of the left
truncated  beta   distribution
when the number of bins, $n$, is 12,
a= 0.12,
b= 1.47,
$\alpha$ = 2.23
and
$\beta$  =3.09.
Vertical and horizontal axes have logarithmic scales.
}
\label{beta_log_ngc2362}
\end{figure*}

\begin{figure*}
\begin{center}
\includegraphics[width=10cm]{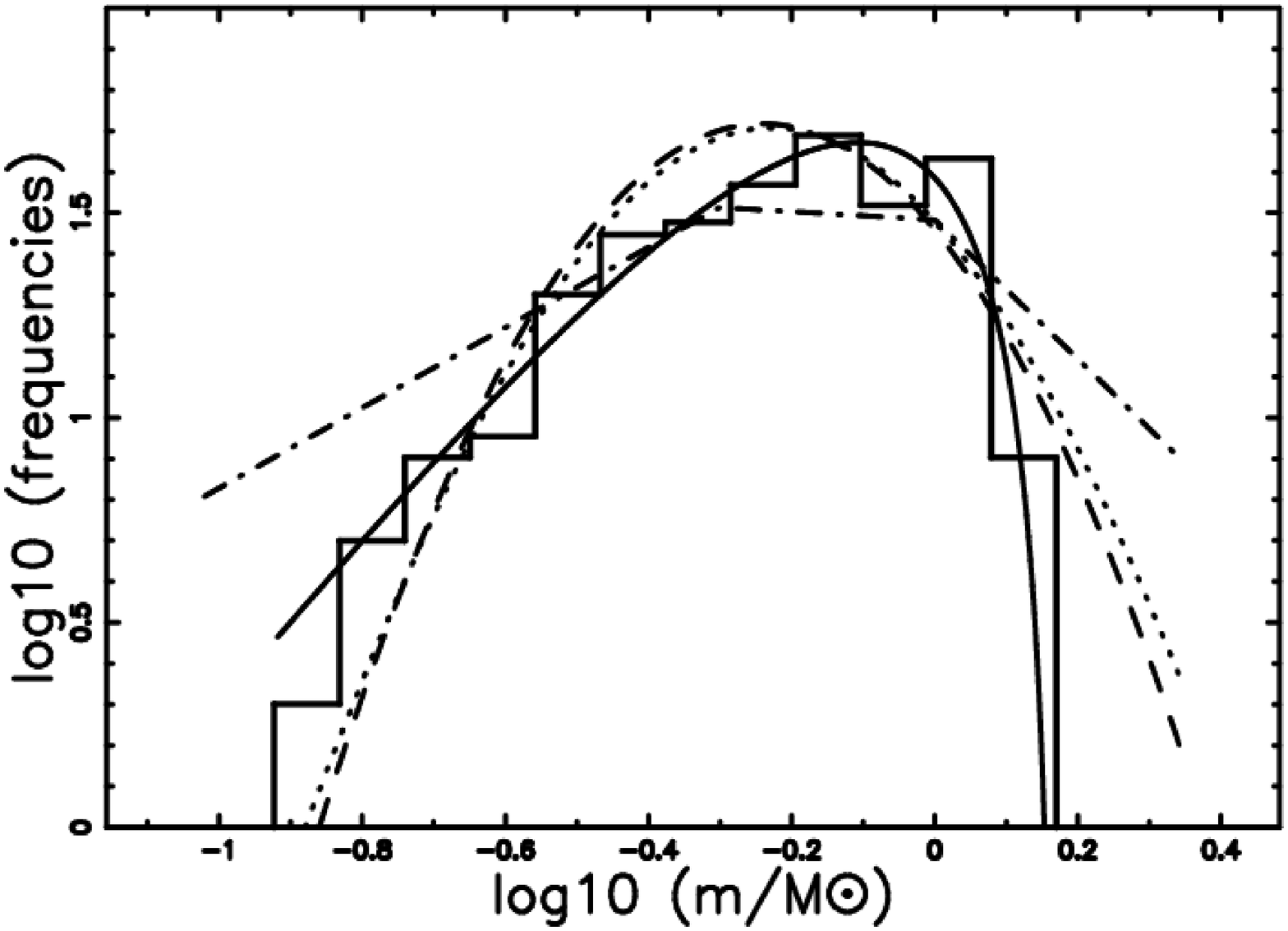}
\end{center}
\caption
{
Histogram (step-diagram)  of  mass distribution
as  given by   NGC 2362 cluster data (272 stars)
with a superposition of the left
truncated  beta   distribution (full line),
the lognornal  (dashed),
the double Pareto lognormal (dotted)
and the four  power laws (dot-dash-dot-dash)  .
Vertical and horizontal axes have logarithmic scales.
}
\label{beta_four_ngc2362}
\end{figure*}

\begin{table}[ht!]
\caption
{
Numerical values of
$\chi_{red}^2$, AIC, probability $Q$,
D, the maximum distance between theoretical and observed DF,
and  $P_{KS}$  , significance level,   in the K-S test
for  the mass distribution
of the NGC 2362 cluster data (272 stars).
The  number of  linear   bins, $n$, is 20.
}
\label{chi2valuesngc2362}
\begin{center}
\begin{tabular}{|c|c|c|c|c|c|c|}
\hline
PDF       &   parameters  &  AIC  & $\chi_{red}^2$ & $Q$  &  D &   $P_{KS}$  \\
\hline
lognormal &  $\sigma$=0.5,$\mu_{LN} =-0.55 $  &  37.64&
1.86    & 0.013   &    0.07305  & 0.10486     \\
\hline
double
&  $\sigma$=0.44 ,$\mu_{LN} =-0.52 $        &
40.42   & 2.02    & 0.008    & 0.066103    &    0.17882       \\
Pareto-lognormal   &   $\alpha$=5, $\beta=5 $ & & & & & \\
\hline
 general~ beta      & $a=0.12 $,$b=$1.47        &
29.09        &    1.31     & 0.17   &  0.059141      &   0.288813    \\
~~~~~~~~            & $\alpha=1.67 $,$\beta$=2.77
&      &  & & &    \\
\hline
 general~ beta     & $a=0.12 $,$b=$1.47        &
31.09       &    1.40     & 0.13   &  0.06412      &  0.20612   \\
+normal (NB)         & $\alpha=1.67 $,$\beta$=2.77, $\sigma=0.001$
&      &  &   & &  \\
\hline
 left       & $a=0.12 $,$b=$1.47        &
31.19      & 1.44  & 0.1  &  0.06158   & 0.24572   \\
truncated~ beta           & $\alpha=2.23 $,$\beta$=3.09
&      &   &    & &  \\
\hline

four    &  Eqn. (\ref{fourpowerngc2362})
& 77.608 & 4.89    & $1.17\,10^{-8} $     & 0.16941   &  $2.60363 \,10^{-7}$ \\
power~ laws  &   
&  ~            &   ~  & & & \\
\hline
\end{tabular}
\end{center}
\end{table}
Figure  \ref{beta_log_ngc2362}  shows the fit
with  the left truncated
beta distribution of  NGC 2362
and Figure  \ref{beta_four_ngc2362}
visually compares the four types of fits for NGC 2362.

\subsection{IMF of  NGC 6611}

\label{secngc6611}
The massive young cluster  NGC 6611 has been carefully analyzed
from the point view of the IMF in  the  range
 $1.5  {M}_{\sun}~>~ {M} \geq  0.02  {M}_{\sun}$.
This  means that also the
BD range is covered, see more details  in \cite{Oliveira2009}
with  data in J/MNRAS/392/1034 at
 the CDS.
Figure  \ref{beta_log_ngc6611}  shows the fit
with  the left truncated
beta distribution of  NGC 6611
and Figure  \ref{beta_four_ngc6611}  shows
a visual comparison of four types of fits for NGC 6611.
Table  \ref{chi2valuesngc6611}  shows
the values of  $\chi_{red}^2$, the AIC, and  the probability $Q$
of  the  astrophysical   fits
and the results
of the K-S test.
Figure  \ref{beta_residuals}
shows the residuals and $\chi^2$ as a function
of the middle value of the logarithmic bin considered,
both for the left truncated  beta and for the lognormal.

\begin{table}[ht!]
\caption
{
Numerical values of
$\chi_{red}^2$, AIC,  probability $Q$,
D, the maximum distance between theoretical and observed DF,
and  $P_{KS}$  , significance level,   in the K-S test
for  the mass distribution
of NGC 6611 cluster data (207 stars + BDs).
The  number of  linear   bins, $n$, is 20.
}
\label{chi2valuesngc6611}
\begin{center}
\begin{tabular}{|c|c|c|c|c|c|c|}
\hline
PDF       &   parameters  &  AIC  & $\chi_{red}^2$ & $Q$
&  D &   $P_{KS}$
\\
\hline
lognormal &  $\sigma$=1.029,$\mu_{LN} =-1.258 $  &  71.24&
3.73    & $1.3\,10^{-7}$  &  0.09366 &  0.04959 \\
\hline
double
&  $\sigma$= 0.979 ,$\mu_{LN} = -1.208 $        &
70.3
 & 3.89    & $ 2.13\,10^{-7} $   &  0.07995      & 0.13523  \\
Pareto-lognormal    &   $\alpha$=4, $\beta=4$
&  ~            &   ~  & & & \\
\hline
 general~ beta      & $a=0.019 $,$b=$1.46        &
39.29      &    1.956     & 0.0123 &  0.11456      & 0.007924 \\
~~~~~~~~            & $\alpha=0.56 $,$\beta$=1.55
&      &  &   & &  \\
\hline
 general~ beta    & $a=0.019 $,$b=$1.46        &
41.3      &    2.08     & 0.008  & 0.09476     & 0.04545   \\
+normal (NB)          & $\alpha=0.56 $,$\beta$=1.55,$\sigma=0.001$
&      &   &   & &   \\
\hline
 left       & $a=0.019 $,$b=$1.46        &
42.09      & 2.13  & 0.005 &  0.06839      &  0.27781  \\
truncated~ beta             & $\alpha=0.55 $,$\beta$=1.6
&      &   &    & &  \\
\hline
four 
    &  Eqn. (\ref{fourpowerngc6611}) 
& 81.39  & 5.18   &  $2.41\,19^{-9}$   & 0.12514     &  $ 2.7239\,10^{-3}$  \\
power~ laws  &   
&  ~            &   ~  & & & \\
\hline
\end{tabular}
\end{center}
\end{table}

\section{Conclusions}

{\bf Motivations}
In the last  50 years,  the IMF has been modeled progressively
by one power law, by two power laws, three power  laws,
and four power laws.
The  three  power law  distribution
has seven parameters
and the four power law
has nine, and they  both have a finite range of existence.
A second widely used fitting function  is the lognormal,
which is characterized by two parameters
and  is defined on the interval $[0, \infty]$.
In this paper, we have described  a  left  truncated
beta PDF which  has
(i)   a lower and an upper  bound,
(ii)  a  finite value  of  probability on the
       lower bound rather than zero,
(iii) two parameters, $\alpha$ and $\beta$, which
       fix  the shape of the distribution,
(iv)    an analytical  expression  for the average value.
Two physical  meanings are  distinguished:
(i)  the upper limit  of the left truncated beta
is  connected  with the maximum stellar  mass,
which  is  $\approx 60 \sunmass$,
(ii) the lower limit is  connected
with an unknown physical  mechanism
which  limits the  distribution in masses.
Further on we remember that the lognormal PDF 
has the important disadvantage of missing the 
well-accepted Salpeter-type high-mass power law.  
The  high masses  behavior   of the various  
PDFs  here analyzed   is   reported 
in Figure \ref{beta_salpeter} for NGC 6611  where  the 
Pareto  and truncated Pareto   PDFs  are evaluated  
for ${M} \geq  0.43  {M}_{\sun}$,
which means  Salpeter slope -2.3 .
From  the previous Figure  the discrepancy  
of the lognormal  and 
double Pareto-lognormal at high masses is evident.

{\bf Goodness of fit tests} The statistical tests here performed
are split in two: (i) a first test  requires binning the data in
order to evaluate $\chi^2$, and  the indicators are
$\chi_{red}^2$, the AIC, and the  probability $Q$; (ii) the K-S
test does not require binning the data, and  the two  indicators
are $D$ and   $P_{KS}$ . These two tests, when applied to  NGC2362
and NGC6611, indicate that the beta family (general and left
truncated) performs better than the lognormal distribution both
when the binning of the data is computed, see Tables
example, the K-S test for the mass distribution of NGC6611
indicates a confidence level of $27\%$ for the left truncated beta
and  $5\%$ for the lognormal. 
New confidence levels  can be found
with the  Anderson-Darling test which is a modification 
of the K-S
test, see \cite{Stephens1974} and the discussion at
https://asaip.psu.edu/Articles/beware-the-kolmogorov-smirnov-test.
Currently   tables of critical values for the  Anderson-Darling
test are available for the lognormal PDF but the critical values
for other PDFs here explored are not yet available, see
http://www.itl.nist.gov/div898/handbook .

{\bf Convolution}
The random sum (convolution) of a general
beta and a normal random variable, as represented by
Equation  (\ref{convolution}),
when applied to  NGC6611
 introduces
a further parameter, $\sigma$,
which  increases $\chi^2_{red}$  and the AIC  from that of
the general beta,
but  decreases   $D$ and   $P_{KS}$  in the  K-S test, see Table
\ref{chi2valuesngc6611}.

\begin{figure*}
\begin{center}
\includegraphics[width=10cm]{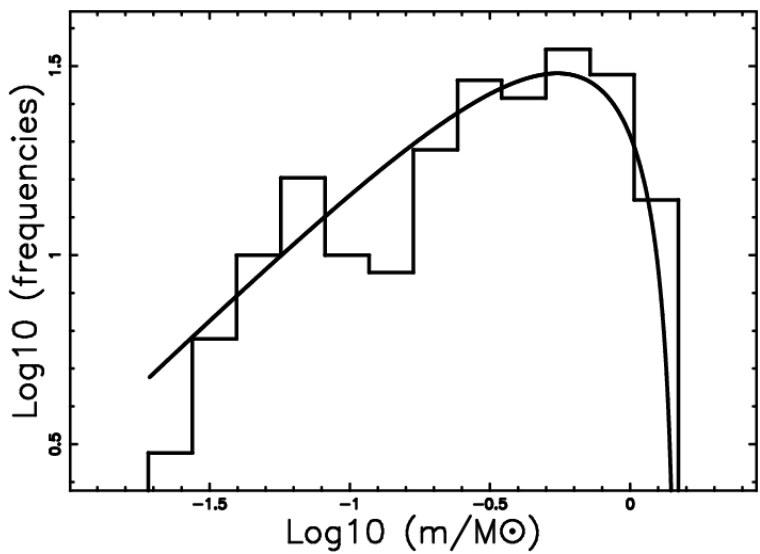}
\end{center}
\caption
{
Logarithmic histogram   of  mass distribution
as  given by   NGC 6611 cluster data (207 stars + BDs)
with a superposition of the left
truncated  beta   distribution
when the number of bins, $n$, is 12,
a= 0.019,
b= 1.46,
$\alpha$ = 0.55
and
$\beta$  =1.6.
Vertical and horizontal axes have logarithmic scales.
}
\label{beta_log_ngc6611}
\end{figure*}

\begin{figure*}
\begin{center}
\includegraphics[width=10cm]{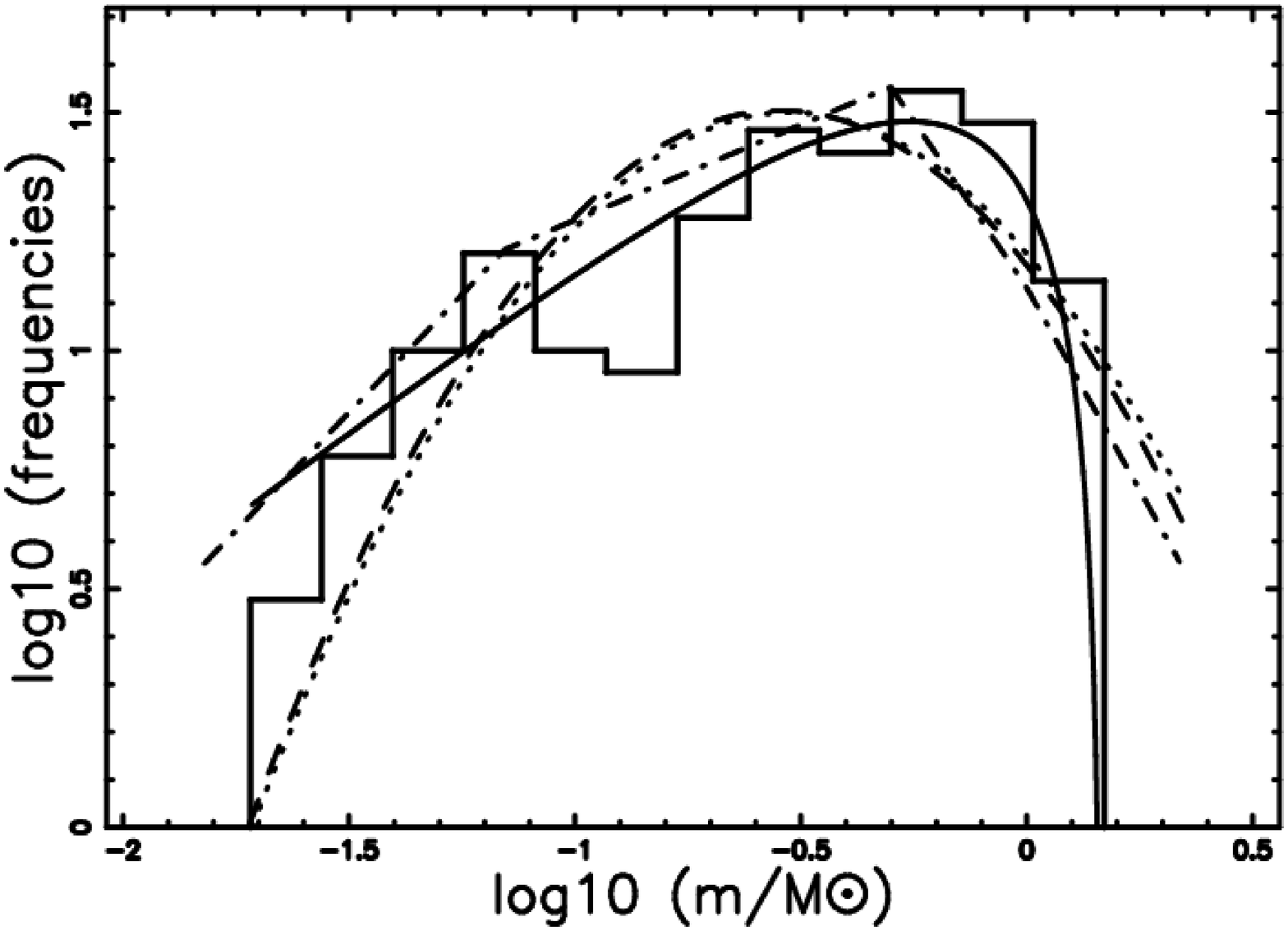}
\end{center}
\caption
{
Histogram (step-diagram)  of  mass distribution
as  given by   NGC 6611 cluster data (207 stars + BDs)
with a superposition of the left
truncated  beta   distribution (full line),
 the lognormal  (dashed),
 the double Pareto lognormal (dotted)
and the four  power laws (dot-dash-dot-dash).
Vertical and horizontal axes have logarithmic scales.
}
\label{beta_four_ngc6611}
\end{figure*}

\begin{figure*}
\begin{center}
\includegraphics[width=10cm]{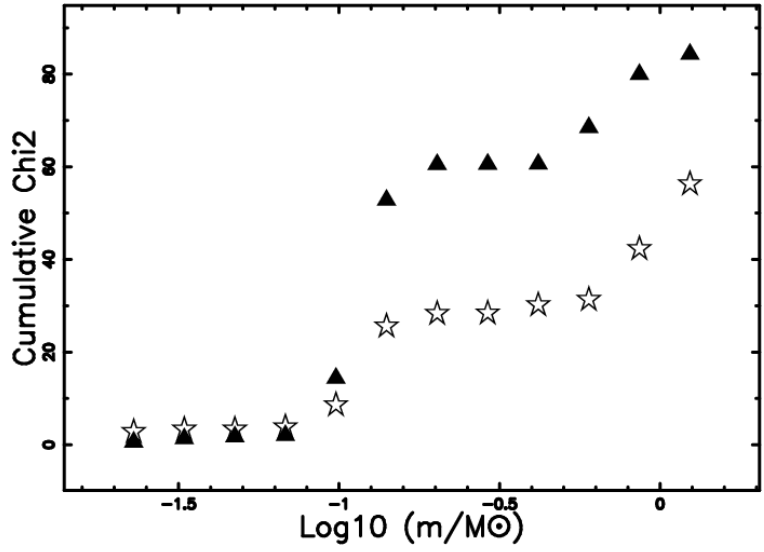}
\end{center}
\caption
{
The residuals of  the fits to NGC 6611 cluster data
when 12 logarithmic bins are considered.
 The empty stars represent  the left truncated beta  PDF
and the filled triangles the lognormal PDF.
}
\label{beta_residuals}
\end{figure*}

\begin{figure*}
\begin{center}
\includegraphics[width=10cm]{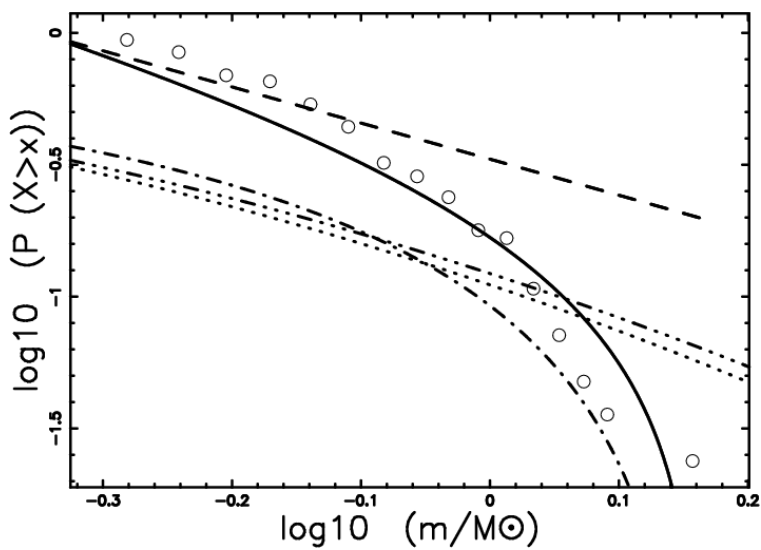}
\end{center}
\caption
{
Survival function
of NGC 6611 cluster data
as  $log_{10}-log_{10}$  plot
 when ${M} \geq  0.43  {M}_{\sun}$  :
data (empty circles),
survival function  of the truncated Pareto  pdf (full line)
(a=0.43,b=1.46,c=1.3)  and
survival function  of the           Pareto  pdf (dashed line)
(c=1.3 , Salpeter slope -2.3).
The left
truncated  beta   distribution (dot-dash-dot-dash)  ,
the lognormal (dotted)  and  the
Double Pareto-lognormal ( dash-dot-dot-dot)
cover all the range in mass with
parameters as in Table \ref{chi2valuesngc6611}.
}
\label{beta_salpeter}
\end{figure*}

{\bf The mode}
A careful attention  should be paid to the
falloff of the IMF towards the brown dwarfs.
The left truncated beta PDF, see  PDF (\ref{betatruncated}),
once  the  numbers of the open cluster NGC6611
are inserted,  see Figure \ref{beta_log_ngc6611},
decreases after  the maximum at
$m  \approx  0.019 \sunmass$.
This fact can be explained  by the following
Taylor expansion
\begin{eqnarray}
f_T(x;0.019,1.46,0.715,2.185) = \nonumber \\
 2.65- 21.64\, \left( x- 0.039 \right) + 336.35\,
 \left( x- 0.039 \right) ^{2}+O \left(  \left( x- 0.039 \right) ^{3}
 \right)
\quad.
\end{eqnarray}
The previous  decreasing function converts itself
into an increasing function when the integration is performed
\begin{eqnarray}
\int_{0.019}^x f_T(x;0.019,1.46,0.715,2.185)
= \nonumber \\
3.50\,x- 10.82\,{x}^{2}+ 112.11\, \left( x-
 0.039 \right) ^{3}
\quad,
\end{eqnarray}
and we recall that the evaluation
of the frequencies
corresponds to an integration.

{\bf Lognormal family}
The recently formulated  double Pareto-lognormal distribution
draws attention to a possible  alternative
to the lognormal.
Our   tests show that
the double Pareto-lognormal
lowers  the  value 
of the maximum  distance, $D$, 
of the K-S test, 
see Tables
\ref{chi2valuesngc2362}  and  \ref{chi2valuesngc6611}.
Inconveniently,  at the moment of writing
there are no analytical evaluations of
the four parameters which characterize  the
double Pareto-lognormal.

{\bf The astronomical sample}
The new PDFS here presented can be tested on an 
astronomical sample
representative  of the IMF.
Currently not all the various catalogs available 
on CDS report the column
of the mass. As an example the promising IMF of  IC 348  , see Figure 11 in
\cite{Oliveira2012} , is not available on CDS.


\newpage

\end{document}